\documentclass[twocolumn,english,aps,prl]{revtex4}
\usepackage[T1]{fontenc}
\usepackage[latin1]{inputenc}
\usepackage{babel}
\usepackage{graphics}

\makeatletter

\providecommand{\LyX}{L\kern-.1667em\lower.25em\hbox{Y}\kern-.125emX\@}

\makeatother
\begin{document}

\title{Nonclassical paths in the recurrence spectrum of diamagnetic atoms}

\author{Brian E. Granger}

\affiliation{ITAMP, Harvard-Smithsonian Center for Astrophysics, 60 Garden Street,
Cambridge, MA 02138}

\author{Chris H. Greene}

\affiliation{JILA and the Department of Physics, University of Colorado, Boulder,
CO 80309-0440}

\begin{abstract}
Using time-independent scattering matrices, we study how the effects
of nonclassical paths on the recurrence spectra of diamagnetic atoms
can be extracted from purely quantal calculations. This study reveals
an intimate relationship between two types of nonclassical paths:
exotic ghost orbits and diffractive orbits. This relationship proves
to be a previously unrecognized reason for the success of semiclassical
theories, like closed-orbit theory, and permits a comprehensive reformulation
of the semiclassical theory that elucidates its convergence properties. 
\end{abstract}
\maketitle

\section{Introduction}

Recurrence spectroscopy \cite{Holle} provides an excellent window
into atomic systems whose classical counterparts are chaotic, such
as atoms in strong magnetic fields. While the energy domain photoabsorption
spectra of these systems are incredibly complex \cite{Iu1991a}, the
time domain physics is relatively simple, at least for short times.
Recurrence spectroscopy gives access to this simple physics by revealing
sharp (recurrence) peaks in the Fourier transform of the photoabsorption
spectra. Semiclassical closed-orbit theory \cite{DuDelos} provides
an elegant interpretation of these short time features \cite{AlLaithy1986}
in the spectra: each recurrence peak is associated with a classical
trajectory in which the electron is launched from and returns to the
nucleus radially. When the properties of these ``closed orbits'' (periods,
stabilities, and so on) are used in the context of semiclassical approximations,
the recurrence spectrum of hydrogen in a strong magnetic field can
be predicted accurately \cite{DuDelos,Main1994a,Karremans1999c}.

\newcommand{\score}{\underline{S}^{\textrm{core}}}

\newcommand{\slr}{\underline{S}^{\textrm{LR}}}

\newcommand{\fcore}[1]{\underline{#1 }^{\textrm{core}}}

\newcommand{\flr}[1]{\underline{#1 }^{\textrm{LR}}}

\newcommand{\ora}[1]{\protect\overrightarrow{#1 }}

\newcommand{\psiphys}[1]{\protect\overrightarrow{\psi }_{#1 }}

\newcommand{\mat}[1]{\underline{#1 }}

\newcommand{\fplus}{\underline{f}^{+}}

\newcommand{\fminus}{\underline{f}^{-}}

\newcommand{\fplusp}{\underline{f}^{+\prime }}

\newcommand{\fminusp}{\underline{f}^{-\prime }}

\newcommand{\fracpar}[2]{\frac{\partial #1 }{\partial #2 }}

There are, however, a number of limitations of closed-orbit theory.
In its traditional formulation, it neglects quantum mechanical (Feynman)
paths that have no classical counterparts. Examples of such paths
are ghost orbits \cite{ghosts} (orbits that have not yet bifurcated
into classical existence) and diffractive orbits \cite{Dando1996a}
(orbits or paths that form when the electron scatters from features
smaller than its wavelength). Clearly, fully quantum mechanical methods
do not suffer from these shortcomings. A number of large scale quantum
calculations have been performed on single channel atoms in strong
magnetic fields \cite{quantum}. While these calculations show excellent
agreement with experiments they provide limited physical insight because
they yield only the total photoabsorption or recurrence cross section,
and do not necessarily help in its interpretation. 

In this paper we outline a method for analyzing recurrence spectra
that combines the interpretive strengths of closed-orbit theory with
the ability of fully quantum mechanical methods to accurately describe
nonclassical effects. This development relies on an exact quantum
mechanical framework for treating atomic photoabsorption in external
fields, which we derived in a previous paper \cite{Granger2000a}.
When accurate quantum calculations are used in this framework, insight
can be gained that is not available in either semiclassical methods
or previous quantum calculations. To illustrate the power of this
approach, we revisit the recurrence spectrum of diamagnetic hydrogen
and study recurrences associated with diffractive and ghost orbits.
Most significantly, an intimate relationship between these two types
of orbits emerges. We conclude by discussing the nontrivial implications
of this discovery for semiclassical theories of photoabsorption. Following
others \cite{Main1994a}, we use the scaled variables (in atomic units)
\( w=2\pi B^{-1/3} \) and \( \epsilon =EB^{-2/3} \), so that in
this paper the energy-like variable is the scaled field \( w \) and
the time-like variable is the scaled action \( \tilde{S} \). 

These advances are enabled by our use of time-independent scattering
matrices to describe the Rydberg electron's motion exactly. When an
external field is applied to an atom there are two such scattering
matrices. First, the core-region \( S \)-matrix \( \score  \) of
quantum defect theory \cite{Seaton1983} is used to characterize the
scattering of the electron off the ionic core at small distances (\( r<10 \)
a.u.). For single channel atoms, such as the alkali-metal atoms, \( \score  \)
is given in terms of the quantum defects \( \mu _{l} \): \( S_{ll^{\prime }}^{\textrm{core}}=\delta _{ll^{\prime }}e^{2\pi i\mu _{l}}. \)

Likewise, the long-range scattering matrix, \( \slr  \), characterizes
the Rydberg electron's motion in the combined Coulomb and diamagnetic
potentials at distances far (\( r>10 \) a.u.) from the ionic core.
As closed-orbit theory first showed, it is both appropriate and fruitful
to use semiclassical approximations for the electronic motion in this
region. The semiclassical approximation to \( \slr  \) (which we
do not utilize in any of the calculations in this paper),\begin{equation}
\label{eq:slrscl}
S^{\textrm{LR},scl}_{ll^{\prime }}(w)=\sum _{j\in co}A^{j}_{ll^{\prime }}(w)e^{i\tilde{S}_{j}w},
\end{equation}
is derived in \cite{Granger2000a}. Like closed-orbit theory, this
involves a sum over the classical closed orbits having scaled actions
\( \tilde{S}_{j} \). The matrix \( \underline{A}^{j}(w) \) varies
slowly with \( w \) and contains other information about the closed
orbits (Maslov index, stability, and so on). This form of the approximate,
semiclassical \( \underline{S}^{\textrm{LR}} \) suggests that the
exact quantum mechanical \( \underline{S}^{\textrm{LR}} \) can be
written in a similar form %
\footnote{The exact \( \protect \underline{S}^{\textrm{LR}} \) can be cast
in the same form as the semiclassical approximation, Eq.~\ref{eq:slrscl},
using a path integral formulation of the scattering amplitudes.
}, but with the sum being over all Feynman paths of the Rydberg electron
that originate near the nucleus with angular momentum \( l^{\prime }, \)
and return to the nucleus with angular momentum \( l \) after scattering
from the long range Coulomb and magnetic fields. While some of these
quantum paths have classical analogues (closed orbits) also appearing
in the semiclassical theory, others are purely quantal paths that
are neglected in semiclassical theories. 

The physics contained in the \( S \)-matrices can be probed with
the photoabsorption cross section and its Fourier transform. The scattering
matrices determine the accurate quantal photoabsorption cross section
\( \sigma (w) \) in a straightforward manner \cite{Granger2000a}:\begin{eqnarray}
\sigma (w) & = & 4\pi ^{2}\alpha \, \textrm{Re}\, \omega _{0}\, \vec{d}\left[ \mat{1}-\score \slr (w)\right] ^{-1}\times \nonumber \\
 &  & \left[ \underline{1}+\score \slr (w)\right] \vec{d}^{\dagger }.\label{eq:finalsigma} 
\end{eqnarray}
In this equation, \( \omega _{0} \) is the frequency of the light
used to excite the transition and \( \vec{d} \) is the atomic dipole
vector. Equation \ref{eq:finalsigma} gives the energy smoothed cross
section according to the following prescription: when the long-range
\( S \)-matrix \( \slr  \) is calculated \cite{Granger2001a} at
a complex value of the scaled field \( w+i\frac{\Gamma }{2} \), Eq.~\ref{eq:finalsigma}
gives the total cross section {}``preconvolved'' with a Lorentzian
of width \( \Gamma  \) \cite{Robicheaux1993a}.

The physical insight contained in the photoabsorption cross section
can be extracted by expanding the matrix \( \left[ \underline{1}-\underline{S}^{\textrm{core}}\underline{S}^{\textrm{LR}}\right] ^{-1} \)
as a geometric series: \begin{equation}
\label{eq:crossexp}
\sigma (w)\sim \textrm{Re }\vec{d}\left[ \underline{1}+2\sum _{n=1}^{\infty }\left( \underline{S}^{\textrm{core}}\mat{S}^{\textrm{LR}}\right) ^{n}\right] \vec{d}^{\dagger }.
\end{equation}
Because the scattering matrices are simply quantum-mechanical amplitudes
to scatter through a region of space a single time, the cross section
can be viewed as an infinite sum over quantum mechanical paths where
the Rydberg electron scatters one \( \left( \underline{S}^{\textrm{core}}\underline{S}^{\textrm{LR}}\right)  \)
or more \( \left( \underline{S}^{\textrm{core}}\mat{S}^{\textrm{LR}}\right) ^{n} \)
times from the core and long-range regions. 

While the semiclassical \( \underline{S}^{\textrm{LR},scl} \) (\ref{eq:slrscl})
can be used in (\ref{eq:crossexp}) to derive a semiclassical theory
\cite{Granger2000a} of photoabsorption that resembles closed-orbit
theory, we emphasize that it is not necessary to make a semiclassical
approximation. In fact, by using an accurate, fully quantum mechanical
\( \underline{S}^{\textrm{LR}} \), detailed information about the
recurrence spectrum can be extracted that is not available in either
closed-orbit theory or other fully quantum mechanical methods. More
specifically, we assert that the study of the Fourier transforms of
individual terms in the expansion of the cross section (\ref{eq:crossexp})
reveals recurrences that appear only at a given order \( (n) \) of
scattering. This ability to study only particular parts of the recurrence
spectrum without making any semiclassical approximation is useful
for studying recurrences associated with nonclassical paths. To illustrate
how this analysis works, we now use it to study nonclassical paths
in diamagnetic hydrogen without performing the detailed analysis required
to repair the semiclassical approach \cite{ghosts}. 

\begin{figure}
{\centering \resizebox*{1\columnwidth}{0.2\textheight}{\includegraphics{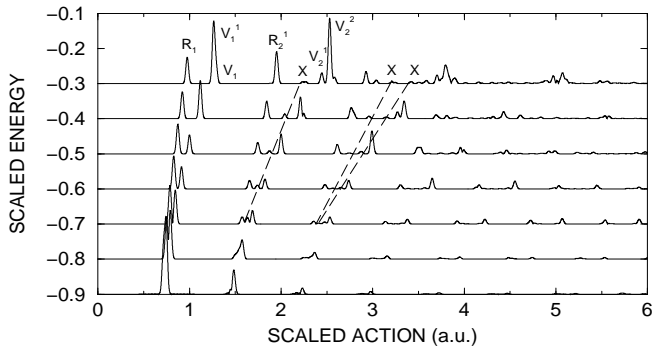}} \par}

\caption{\label{fig:first}The Fourier transform, or recurrence strength,
of the linear term in the expansion of the photoabsorption cross section
\protect\( 2Re\protect \vec{d}\protect \underline{S}^{\textrm{LR}}(w)\protect \vec{d}^{\protect \dagger }\protect \)
is plotted for hydrogen (even parity, \protect\( m=0\protect \))
at seven scaled energies (\protect\( \protect \epsilon =-0.9\protect \rightarrow -0.3\protect \)).
Recurrence peaks corresponding to primitive closed orbits are labeled
using the notation of \cite{Holle}. Because an accurate, fully quantum
mechanical \protect\( \protect \underline{S}^{\textrm{LR}}(w)\protect \)
has been used, recurrence peaks associated with ghost orbits can also
be seen. The recurrence peaks indicated by dashed lines are ghost
orbits (\protect\( X's\protect \)) \cite{ghosts} that bifurcate
into real closed orbits through saddle-node bifurcations at higher
scaled energies than shown here. }
\end{figure}

\begin{figure}
{\centering \resizebox*{1\columnwidth}{0.2\textheight}{\includegraphics{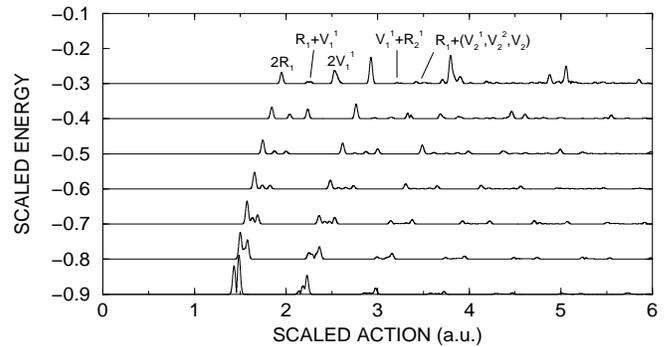}} \par}

\caption{\label{fig:second}The recurrence strength, of the quadratic term
in the expansion of the photoabsorption cross section \protect\( 2Re\protect \vec{d}\left( \protect \underline{S}^{\textrm{core}}\protect \underline{S}^{\textrm{LR}}(w)\right) ^{2}\protect \vec{d}^{\protect \dagger }\protect \)
is plotted for diamagnetic hydrogen (see Fig.~\ref{fig:first}).
No semiclassical approximations have been made, which results the
appearance of nonclassical recurrences that are related to the electron
scattering twice from the long range region. More specifically, recurrences
associated with combinations (\protect\( R_{1}+V_{1}^{1},\ldots \protect \))
of two closed orbits are seen, which are classically forbidden because
of the strong backscattering of classical low-\protect\( l\protect \)
Coulomb scattering. Classically-allowed double repetitions of the
primitive closed orbits (\protect\( 2R_{1},2V_{1}^{1},\ldots \protect \))
are also seen. }
\end{figure}

\begin{figure}
{\centering \resizebox*{1\columnwidth}{0.2\textheight}{\includegraphics{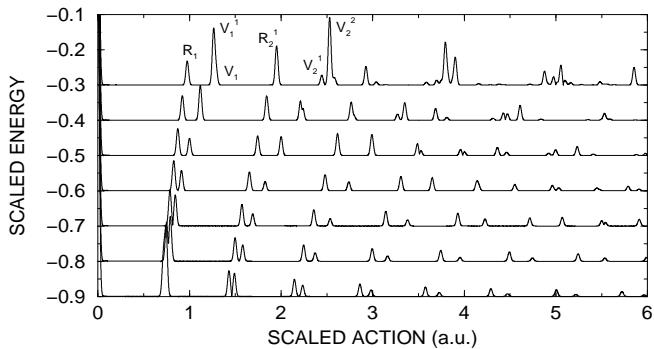}} \par}

\caption{\label{fig:full}The recurrence strength of the total photoabsorption
cross section \protect\( \protect \sigma (w)\protect \), Eq.~(\ref{eq:finalsigma}),
is plotted for diamagnetic hydrogen (see Figs.~\ref{fig:first} and
\ref{fig:second}). While all nonclassical effects (ghost orbits,
combination orbits) are included exactly in this purely quantal calculation,
neither the ghost orbits (Fig.~\ref{fig:first}) nor the combination
orbits (Fig.~\ref{fig:second}) survive in the total cross section.
They cancel accurately.}
\end{figure}

There are two main types of nonclassical paths in diamagnetic atoms:
ghost orbits and diffractive orbits. Because ghost orbits and diffractive
orbits appear at different orders in the expansion of the photoabsorption
cross section, their contributions can be isolated. This analysis
is presented in Figs.~\ref{fig:first}, \ref{fig:second} and \ref{fig:full}
where we present accurate quantum calculations of the scaled recurrence
spectrum of diamagnetic hydrogen. These results were obtained by using
an \( \protect \underline{S}^{\textrm{LR}}(w) \) calculated with
a \( B \)-spline \( R \)-matrix technique and \( \underline{S}^{\textrm{core}}=\underline{1} \)
in the various terms of Eq.~\ref{eq:crossexp} \cite{Granger2001a}.
The dipole vector \( \vec{d} \) from the \( 2p \), \( m=0 \) state
to the even parity \( m=0 \) final states shown here is given in
\cite{DuDelos}. In the scaled variables used here the energy-like
variable, \( w \), is related to an effective Planck's constant \( \hbar _{eff}=2\pi /w \).
With this in mind, we study the Fourier transform of the cross section
\( \sigma (w) \) at relatively low values of \( w \) (\( 100-500 \))
so that nonclassical effects are more pronounced. 

The first order recurrence spectrum \( \left| \textrm{FT}\left[ 2Re\vec{d}\underline{S}^{\textrm{LR}}(w)\vec{d}^{\dagger }\right] \right| , \)
pictured in Fig.~\ref{fig:first}, shows recurrence peaks from only
primitive closed orbits and ghost orbits. To identify which recurrence
peaks correspond to ghost orbits, we first perform a classical calculation
of all the relevant closed-orbits at the scaled energy of interest.
Then, when a first order recurrence peak is seen at a scaled action
where no closed orbit exists, a ghost orbit has been found. More specifically,
as Fig.~\ref{fig:first} shows, we can readily identify the so called
{}``exotic'' ghost orbits that bifurcate into real closed orbits
through a saddle-node bifurcation at (much) higher scaled energies.
A semiclassical theory for these ghosts has been worked out by Main
and Wunner \cite{ghosts} who predict an exponential falloff of the
ghost amplitude at scaled energies below the classical bifurcation
point. The first order recurrence spectrum in Fig.~\ref{fig:first}
clearly does not show such a decay of the ghost orbit recurrences.
This apparent discrepancy, and its resolution, will be discussed below,
after the second order recurrence spectrum has been presented.

The Fourier transform of the second order term in the expansion of
the cross section, \( \left| \textrm{FT}\left[ 2Re\vec{d}\underline{S}^{\textrm{LR}}(w)\underline{S}^{\textrm{LR}}(w)\vec{d}^{\dagger }\right] \right| , \)
is pictured in Fig.~\ref{fig:second} and shows only recurrences
due to double repetitions (\( 2R_{1},2V_{1}^{1},\ldots  \)) of primitive
closed orbits and double combinations (\( R_{1}+V_{1},\ldots  \))
of different closed orbits. While the presence of recurrences due
to repetitions of the same closed orbit are predicted in semiclassical
theories, the presence of combinations of \emph{different} orbits
in hydrogen is surprising from a semiclassical perspective. Typically
it is argued (see \cite{Courtney1996a}, p.185) that because scattering
of a low-\( l \) state by a pure Coulomb potential is strongly backscattered,
the quantum wave returning to the nucleus along a given closed orbit
can only scatter back out into the same closed orbit and not into
any other closed orbits. However, we emphasize that this picture is
strictly true only in the classical limit (\( \hbar _{eff}=0 \)).
At any finite value of \( \hbar _{eff} \), the \( \Delta \theta \Delta l\gtrsim \hbar  \)
uncertainty relation limits the validity of this highly directional
classical scattering argument. The accurate quantum calculations presented
in Fig.~\ref{fig:second} support this claim by showing second-order
recurrences that appear at the scaled actions of combinations of primitive
closed orbits.

\begin{figure}
{\centering \resizebox*{1\columnwidth}{0.13\textheight}{\includegraphics{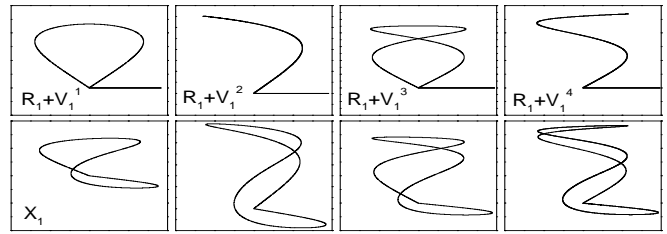}} \par}

\caption{\label{fig:trajs}This figure shows the topological similarity between
the exotic ghost orbits (bottom) and their combination orbit counterparts
(top). The orbits are plotted in cylindrical (\protect\( \rho ,z\protect \))
coordinates at a scaled energy slightly above that where the ghost
orbit has bifurcated into existence (as a true closed orbit) through
a saddle-node bifurcation. }
\end{figure}

The main point of the present letter is based on a simple observation
about the first and second order recurrence spectra shown here. A
quick comparison of Figs.~\ref{fig:first} and \ref{fig:second}
reveals an interesting trend about the recurrences of (exotic) ghost
orbits and diffractive combination orbits. Namely, they have similar,
if not identical, scaled actions and recurrence amplitudes. This correspondence
holds for every case that we have studied, which suggests that there
is a one to one relationship between exotic ghost orbits and diffractive
combination orbits. This proposal is further elucidated by the topologies
of the trajectories of the ghost and combination orbits, which are
plotted in Fig.~\ref{fig:trajs}. \textsl{In all cases, orbits having
similar scaled actions are also topologically similar.} We now turn
to the most interesting aspect of this relationship, which emerges
when the total recurrence spectra (all orders of scattering) are plotted.

Surprisingly, the total recurrence spectra, \( \left| \textrm{FT}\left[ \sigma (w)\right] \right|  \),
shows neither exotic ghost orbits, nor combination orbits. More specifically,
the ghost orbit recurrences (first order) cancel with the combination
orbit recurrences (second order) in the total recurrence spectrum
for hydrogen photoabsorption. Again, except very near the classical
bifurcation points, this cancellation seems universal, and essentially
complete.

This relationship between exotic ghost orbits and combination orbits
in hydrogen can be expressed formally when the exact quantum \( \underline{S}^{\textrm{LR}} \)
is written in the form,\begin{equation}
\label{eq:primandghsot}
\mat{S}^{\textrm{LR}}=\sum _{j\in co}\mat{S}_{j}+\sum _{p\in ghosts}\mat{S}_{p},
\end{equation}
 where the first sum is over all paths associated with a classical
closed orbit, and the second sum is over the ghost orbits that are
not associated with a classically-allowed real closed orbit at the
scaled energy of interest. Based on our accurate quantum calculations
presented above, we postulate the following: \textsl{for every exotic
ghost \( p \) there exist two or more primitive closed orbits (\( i,j,k\ldots  \))
such that} \begin{equation}
\label{eq:mathcancel}
\mat{S}_{p}+P(\mat{S}_{i}\mat{S}_{j}\underline{S}_{k}\ldots )\sim e^{-(\epsilon _{b}-\epsilon )},
\end{equation}
\textsl{when the current scaled energy \( \epsilon  \) is less than
the scaled energy \( \epsilon _{b} \) where the ghost orbit bifurcates
into a real closed orbit}. The function \( P() \) in (\ref{eq:mathcancel})
denotes the sum of all possible permutations of the primitive orbit
\( S \)-matrices. The exponentially decaying recurrence amplitude
that is predicted by Main and Wunner actually consists of two pieces:
a diffractive orbit and a ghost orbit. 

We close by outlining the implications of the cancellation (in hydrogen)
between exotic ghost and combination orbits for semiclassical theories.
Most importantly, in atoms other than hydrogen, the interference between
these two types of paths is no longer completely destructive. This
is because the diffractive combination orbits scatter twice from the
ionic core, picking up a phase \( \exp (4i\pi \mu _{l}) \), while
the exotic ghost orbits scatter only once from the ionic core during
their trajectory acquiring a phase of \( \exp (2i\pi \mu _{l}) \).
The resulting nondestructive interference leads to the well know core-scattered
recurrences in the total spectra. 

The cancellation condition, Eq.~(\ref{eq:mathcancel}), can be incorporated
into the photoabsorption cross section, Eq.~(\ref{eq:crossexp}),
to derive a semiclassical theory of photoabsorption for atoms other
than hydrogen. The resulting semiclassical photoabsorption cross section
is \cite{Granger2001a},\begin{equation}
\label{eq:sclsigmaresum1}
\sigma _{scl\sim }\textrm{Re }\vec{d}\left[ \underline{1}+2\score \mat{S}^{\textrm{LR}}_{scl}\sum _{n=0}^{\infty }\left( \mat{T}\mat{S}^{\textrm{LR}}_{scl}\right) ^{n}\right] \vec{d}^{\dagger }
\end{equation}
 which is valid except near the classical bifurcation points. This
is exactly the result that other researchers have derived in extending
closed orbit theory to treat nonhydrogenic atoms \cite{Dando1996a,Shaw1998a},
recast in terms of scattering matrices. Most importantly, this form
of the cross section is a power series in the matrix \( \mat{T}=\mat{1}-\score  \),
which vanishes for hydrogen. We emphasize that the assumptions implicit
in closed-orbit theory, namely the subtle relationship between exotic
ghost orbits and diffractive orbits, have remained obscured until
now. This work elucidates these subtleties and describes how a meaningful
semiclassical approximation emerges out of a careful consideration
of the exact quantum mechanics. 

The main advantage in writing the result of closed orbit theory in
terms of \( S \)-matrices is that it is easily recognizable as a
geometric series, which can be resummed analytically,\begin{equation}
\label{eq:sclsigmaresum2}
\sigma _{scl}\sim \textrm{Re}\vec{d}\left( \mat{1}+2\score \mat{S}^{\textrm{LR}}_{scl}\left[ \mat{1}-\mat{T}\mat{S}^{\textrm{LR}}_{scl}\right] ^{-1}\right) \vec{d}^{\dagger },
\end{equation}
 to include all orders of core-scattering automatically. Furthermore,
the convergence properties of the series, Eq.~\ref{eq:sclsigmaresum1},
are well known: Eq.~(\ref{eq:sclsigmaresum1}) converges absolutely
to Eq.~(\ref{eq:sclsigmaresum2}) when:\begin{equation}
\label{eq:converge}
\left| \left( \score -\mat{1}\right) \mat{S}^{\textrm{LR}}_{scl}\right| <1.
\end{equation}
 While this is a formal requirement that may or may not be satisfied
in a given circumstance, Eqs.~(\ref{eq:sclsigmaresum1}), (\ref{eq:sclsigmaresum2})
and (\ref{eq:converge}) provide a well established framework for
further study of the convergence properties of closed orbit theory.

This work was funded in part by the U.S. Department of Energy, Office
of Science. Discussions with John Delos and John Shaw were helpful
in this work.

\end{document}